\documentclass[twocolumn,showpacs,preprintnumbers,amsmath,amssymb,superscriptaddress]{revtex4}
\usepackage{graphicx}
\usepackage{dcolumn}
\usepackage{bm}
\usepackage{multirow}
\usepackage{placeins}
\usepackage{hyperref}
\begin{document}


\title{The relativistic glider revisited}

\author{L.~Bergamin}
\email{bergamin@tph.tuwien.ac.at}
\affiliation{European Space Agency, The Advanced Concepts Team \\ Keplerlaan 1, 2201 AZ Noordwijk, The Netherlands}

\author{P.~Delva}
\email{Pacome.Delva@esa.int}
\affiliation{European Space Agency, The Advanced Concepts Team \\ Keplerlaan 1, 2201 AZ Noordwijk, The Netherlands}

\author{A.~Hees}
\email{aurelien.hees@oma.be}
\affiliation{European Space Agency, The Advanced Concepts Team \\ Keplerlaan 1, 2201 AZ Noordwijk, The Netherlands}
\affiliation{Observatoire Royal de Belgique (ORB) \\ Avenue Circulaire 3, 1180 Bruxelles, Belgium}

\date{\today}


\begin{abstract}
In this paper we analyze some aspects of the ``relativistic glider'' proposed by Gu\'eron and Mosna more in detail. In particular an explicit weak gravity and low velocity expansion is presented, the influence of different initial conditions are studied and the behavior of the glider over a longer integration time is presented. Our results confirm that the system can be used as a glider, but is not able to stop or even revert the fall of an object.

\end{abstract}
\maketitle

\section{Introduction}

Changing the shape of a falling body in a gravitational field can induce changes in its motion. Such effects are well known and also could be used in space. In Newtonian gravity two body systems (two spacecraft connected by a tether) orbiting the Earth have been studied in \cite{Martinez:1987om,Landis:1991sr,Landis:1992ro}. Once-per-orbit modulations of the tether length \cite{Martinez:1987om,Landis:1992ro} or spinning tether systems with a modulation of the tether length linked to the spinning frequency \cite{Landis:1992ro} may be used as systems for propellentless propulsion.

More recently Wisdom \cite{Wisdom:2003aa} and Gu\'eron et al.\ \cite{Gueron:2005ye,Gueron:2006fq} studied similar situations within general relativity. As most important difference the modification of the motion is no longer a resonant effect, thus the frequency of the change of shape is not linked to an orbital or spinning frequency. As pointed out in \cite{Gueron:2006fq}, their ``gliding effect'' has several advantages over the ``swimming effect'' by Wisdom, in particular it is not suppressed at high frequency and thus it appears to be possible to obtain a relevant displacement even in a very weak gravitational field such as the Earth more easily. However, the work of Gu\'eron et al.\ \cite{Gueron:2006fq} only covers the radial motion and even there just presents the result of the integration over one period of cyclic changes. In this paper we present a more detailed analysis of the effect: it is identified in a weak gravity expansion, the dependence on the initial conditions is studied, it is shown how the system behaves over a longer integration time and finally we discuss different ways of implementing the system constraint.

\section{Model}

We consider the same model as introduced in Ref.~\cite{Gueron:2006fq}: two point masses connected by a massless strut which moves in the gravitational field of Schwarzschild spacetime. Analogously to that work, we use Schwarzschild coordinates as explicit coordinates and the motion is restricted to radial fall, such that the system is described by Schwarschild time $t$ and the radii of the two point masses, $r_i$. The massless strut is implemented as a time dependent constraint between the two radii: $r_1(t)=r_2(t)+l(t)$, where

\begin{equation} \label{CONSTR} l(t;\omega,\alpha,\delta_l)=l_0+\delta_l \ \exp\left[\frac{(1+\alpha-2\omega t)^2 }{(1+\alpha^2)\omega t(-1+\omega t)}\right] \end{equation} 

for $t\in[0,T]$, and then is periodic. The time dependent length of the strut, $l(t)$, is described by four parameters: its frequency $\omega=1/T$, its amplitude $\delta_l$, its minimum length and an asymetry parameter $\alpha$. The asymmetry parameter $\alpha$, taking values in the range $[-1,1]$, indicates how much the constraint fails to be symmetric with $\alpha = 0$ being the symmetric case. In Section \ref{sec:fermi} some issues about the implementation of the constraint are discussed.

Putting the pieces together, the equations of motion are derived by using the action\footnote{we work in geometrical units $G=c=1$.}

\begin{equation}\label{action}
S=-\int dt \left[ \sqrt{L_1} + \sqrt{L_2} +\lambda \left(r_2-r_1-l(t)\right) \right]\ ,
\end{equation}
where $\lambda$ is a Lagrange multiplier enforcing the constraint and 
\begin{equation}
L_i=1-\frac{r_s}{r_i}-\left(1-\frac{r_s}{r_i}\right)^{-1}\dot{r}_i^2\ ,
\end{equation}
with $r_s$ being the Schwarzschild radius. We suppose that the two masses are equal. In the following we also use the geometrical center of the system, $r=(r_1+r_2)/2$. Then the system can be described with the variable $r$ and the constraint $l(t)$.

We are interested in the difference between the vibrating system and a non-vibrating system having the same initial conditions. This radial difference is noted $\delta r(t)=r(t)-r_{0}(t)$ where $r(t)$ is the geometrical centre of the vibrating system and $r_{0}(t)$ is the geometrical centre of the non-vibrating one (our reference motion).

\section{Expansion}
The initial radius of the \emph{apparatus} is $r(t=0)=R$. We first suppose that it is launched with a nul velocity $\dot{r}(t=0)=0$. We introduce the order of magnitude $O_1 = r_s/R \ll 1$. The quantity $(O_1)^{n}$ is denoted $O_n$. We suppose that the amplitude of the constraint is small compare to the Schwarzschild radius, and is of the first order: $\delta_l/r_s \sim O_1$. Then we can expand the equations of motion, derived from the action~(\ref{action}), around a reference motion: $r=r_0+\delta r$, where $r_0$ is the trajectory of the non-vibrating system, and $\delta r \lesssim \delta_l$. We suppose that we are in a weak gravitational field, and in the low velocity limit. Then we also expand the equations of motion with respect to $r_s/r$ and $\dot{r}$. The resulting equations of motion are:

\begin{equation}
\label{motion}
\ddot{\delta r} = - \dfrac{3 \ddot{l}}{4} \left[ \dot{l}  \left( \dot{r_0} + \dot{\delta r} \right) + O_2 \right] \ ; \ \ddot{r}_0 = - \dfrac{1}{r_0} \left( \dfrac{r_s}{2 r_0} + O_2 \right)
\end{equation}

\begin{figure}
 \centering
 \includegraphics[width=8.3cm]{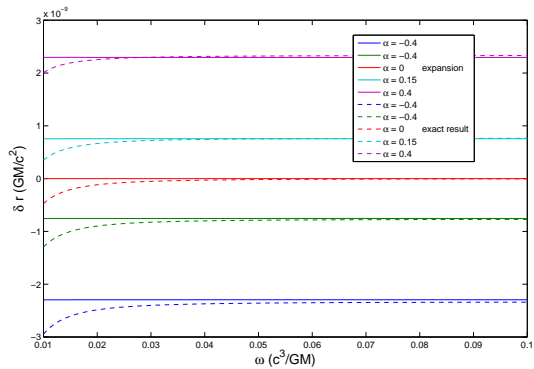}
 \caption{Comparison of the result from Ref.~\cite{Gueron:2006fq} with the expansion presented here, for $R=120$ and $\dot{r}(t=0)=0$.}
 \label{figexpansion}
\end{figure}

The comparison between the exact result given in~\cite{Gueron:2006fq} and the expansion is shown on Fig.~\ref{figexpansion}. The agreement is good for high oscillation frequencies. Indeed, the integration time for each point is $T=1 / \omega$. For low frequencies, the expansion is not valid anymore for two reasons: the distance covered by the system becomes non-negligeable compared to $R$, such that the $r_s/r$ expansion breaks down; the velocity of the system becomes non-negligeable compared to $c$, such that the low velocity limit breaks down. Additionally the validity of the expansion imposes some limitations on the parameters $\omega$ and $\alpha$.

After one period $T$ of oscillation, the velocity $\dot{r_0} \sim g T \sim O_1 / \omega R$, where $g=r_s/R^2$, if the distance covered by the system during the time $T$ is very little compare to $R$: $\dot{r_0} T \ll R$. Therefrom we obtain the condition $\omega R \gg O_{1/2}$, and we get $\dot{r}_0 \ll O_{1/2}$. On the other hand, $\dot{\delta r} \sim \omega \delta r \lesssim \omega \delta_l \sim \omega R O_2$. Supposing that $\omega R \lesssim O_{-3/2}$, we get $\dot{\delta r} \lesssim O_{1/2}$. The $r_s/r$ expansion is valid for $\Delta r_0 \ll R$, where $\Delta r_0$ is the distance covered by the system during the time $T$. We have $\Delta r \sim g T^2 \sim O_1 / \omega^2 R$, thus leading to $\omega R \gg O_{1/2}$, which is coherent with the low velocity approximation. Then the expansion~(\ref{motion}) is valid for

\begin{equation} \label{cond} \dfrac{\sqrt{r_s}}{R^{3/2}} \ll \omega \lesssim \dfrac{\sqrt{R}}{r_s^{3/2}}\ . \end{equation}

In Fig.~\ref{figexpansion} $R=120 \ (GM/c^2)$ has been chosen, a value valid for a compact object such as a black hole or a neutron star. Then $O_1 \simeq 1.7 \cdot 10^{-2}$ and the expansion is valid for $1.1 \cdot 10^{-3} \ll \omega r_s \lesssim 3.9$ (in units of $c^3/GM$). We can see from the condition~(\ref{cond}) that for higher radius the expansion is valid for a broader range of frequencies. For Earth orbit with $O_1 \sim 10^{-9}$, the expansion is valid for $1.6 \cdot 10^{-14} \ll \omega \lesssim 1.6 \cdot 10^{4}$ (in units of $c^3/GM$). 

There is also a condition on $\alpha$: if the assymetry parameter $\alpha \rightarrow 1$, then $\dot{l} \rightarrow 1$ and the low velocity limit is no longer valid. Then the expansion~(\ref{motion}) is not valid when $\alpha$ is too near from one.

\begin{figure}
 \centering
 \includegraphics[width=8.3cm]{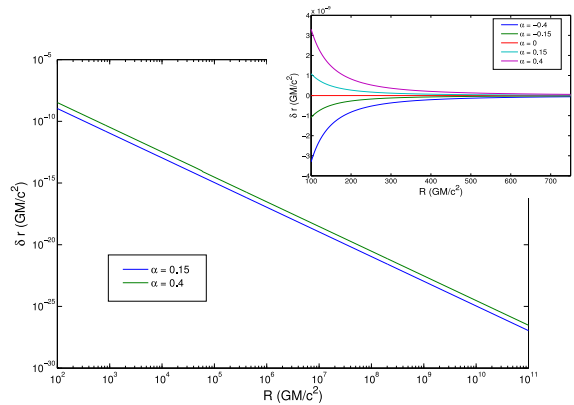}
 \caption{Representation of the radial deviation $\delta r$ (obtained by solving the expanded equation of motion (\ref{motion})) after one period of oscillation for $\omega=0.05$ and $\delta_l=5 \ 10^{-3}$. In the figure below, a log-log scale is used and in the top figure, a normal scale is used. The net displacment $\delta r$ is proportional to $R^{-2}$. This is confirmed here up to $R=10^{11}$.}
 \label{explog}
\end{figure}

In \cite{Gueron:2006fq} the net displacement after one period $\delta r=\delta r(T=1/\omega)$ is estimated to be
\begin{equation}
\label{comp}
\delta r \simeq \Gamma(\alpha) \dfrac{\delta_l}{R^2} \dfrac{G M}{c^2}\ ,
\end{equation}
where $\Gamma$ is a dimensionless parameter that depends on the assymetry parameter $\alpha$. In that work this relation is confirmed numerically for $R$ ranging from $50$ to $10^7$ (in units of $GM / c^2$). Using the expansion derived in this section we can confirm this relation for higher radii up to $R=10^{11}$. This in particular includes orbits around the Earth, where $R \sim 10^9$. In Figure~\ref{explog} the variation of $\delta r$ with repect to $R$ as derived from the expanded equations~(\ref{motion}) are displayed.

\section{Changing the initial conditions}
In Ref.~\cite{Gueron:2006fq} the dependence of the displacement effect on the starting radius $R$ was investigated and a $1/R^2$ was found. However, no other dependence on the initial conditions was considered.

\begin{figure}[htb]
	\centering
		\includegraphics[width=8.3cm]{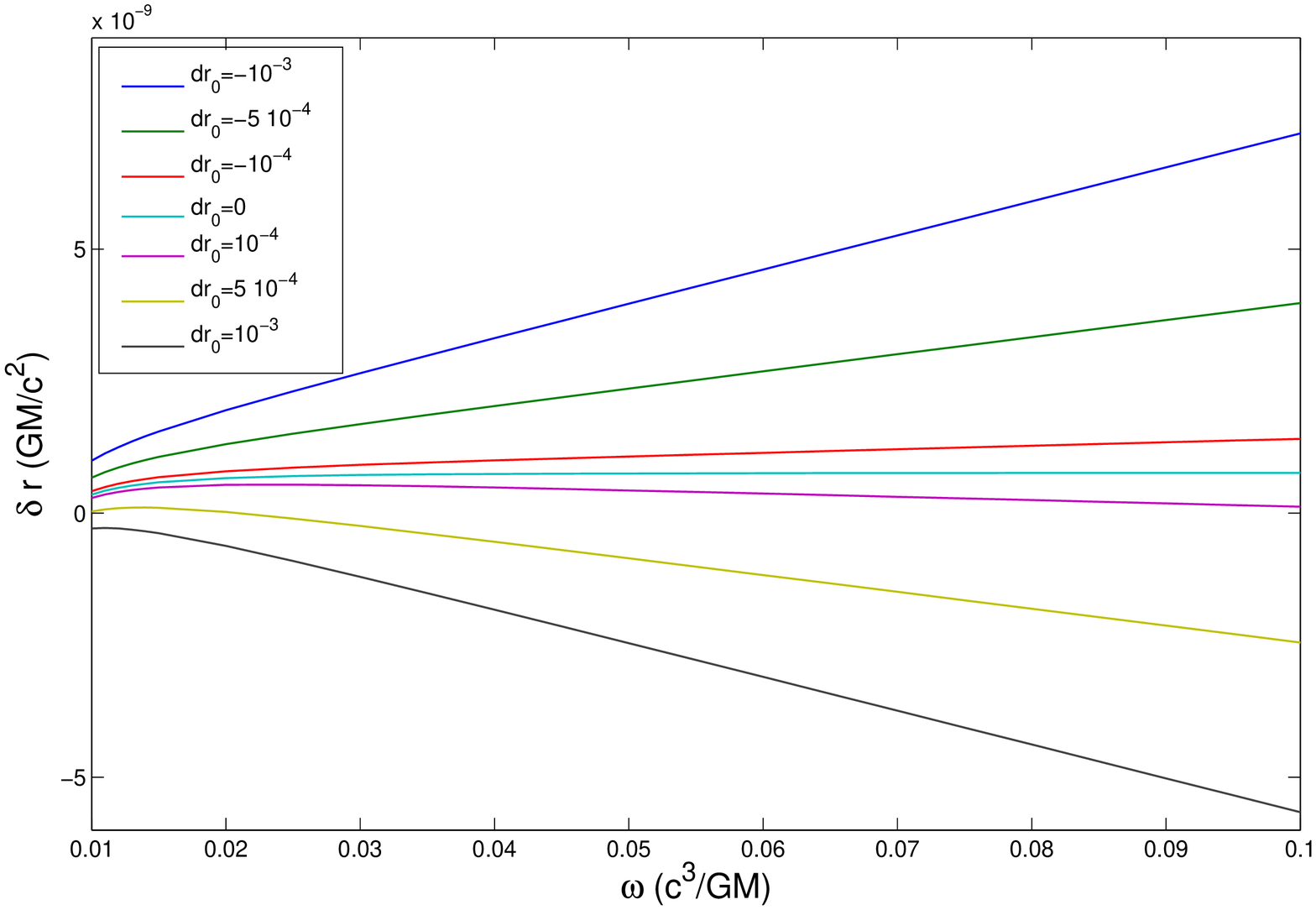}
	\caption{Representation of $\delta r$ as a function of $\omega$ (and for $\alpha=0.15$) for different values of the initial radial velocity (noted $dr_0$ expressed in fraction of $c$). We can see that for a falling system, the effect is increase while for a system going to upwards, the effect is decreased.}
	\label{figInitDr}
\end{figure}

Let us first look at different \emph{radial} initial velocities, $\dot r(t=0)\neq 0$. In Figure \ref{figInitDr}, $\delta r$ for different initial velocities at a positive value of $\alpha$ are presented. As can be seen, the effect increases if the initial velocity is negative (the system already falls down at $t = 0$), while in the opposite case (the system is thrown up at $t = 0$), the effect is decreased and at higher velocity becomes negative. This behavior certainly is not surprising, since the work performed by the oscillations is expected to change sign when the system moves upwards.

\begin{figure}[htb]
	\centering
		\includegraphics[width=8.3cm]{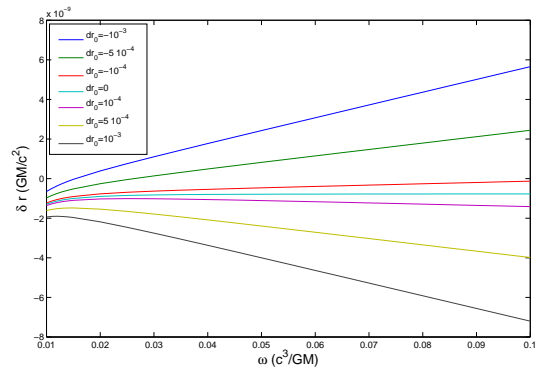}
	\caption{Representation of $\delta r$ as a function of $\omega$ (for $R=120$ and for $\alpha=-0.15$) for different values of the initial radial velocity (noted $dr_0$ expressed in fraction of $c$). For a fastly falling system it is still possible to get a positive effect.}
	\label{figInitDr2}
\end{figure}

Still, changing the sign of the asymmetry parameter for $\dot r (t=0)>0$ is not an option to obtain a positive displacement. This situation is depicted in Figure \ref{figInitDr2}, where different initial velocities with negative values of $\alpha$ are presented. The change in the sign of $\alpha$ induces a vertical shift of all curves. Thus, a positive effect only is achievable for a system falling down with high velocity, while for initial conditions $\dot r = 0$ a negative value results over the whole frequency range, as was already found in \cite{Gueron:2006fq}.

\begin{figure}[htb]
	\centering
		\includegraphics[width=8.3cm]{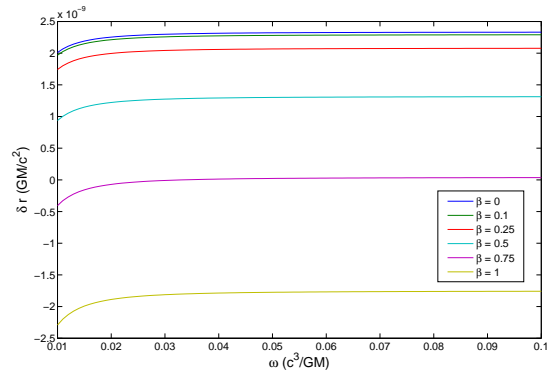}
	\caption{$\delta r$ as a function of $\omega$ for different values of initial angular velocity, expressed as a fraction of the circular angular velocity. In this example $R=120$ and $\alpha=0.4$ has been chosen.}
	\label{figTan}
\end{figure}

A different question concerns a non-trivial \emph{angular} velocity as initial condition. For simplicity we assume that the strut stays aligned in radial direction during the complete motion. Thus it is sufficient to add one degree of freedom, the angle $\varphi$, to the system in order to allow rotations around the central body. This model is described more in detail in Ref.~\cite{Bergamin:2008ar2}.
Figure \ref{figTan} shows the behavior of $\delta r$ for different initial angular velocities. The angular velocities are represented in fractions of the angular velocity of a circular orbit. At $\beta = 0$ the system falls down radially, at $\beta = 1$ the (non-vibrating) system is put into a circular orbit around the central body which means
\begin{equation}
\dot{\varphi}_c^2=\frac{GM}{R^3}\ .
\end{equation}
It can be seen that the effect decreases if the angular velocity is increased and becomes negative at high values. Still, the displacement is not very sensible to the initial conditions as long as the angular velocity remains small. This shows that the vibrations indeed can be used for ``gliding,'' but will not help to increase the energy of an orbit. For further discussions on the effects of vibrating systems in an orbit around a central mass see Ref.~\cite{Bergamin:2008ar2}.
Similar conclusions apply for negative values of $\alpha$, but since for these values $\delta r < 0$ at $\beta = 0$ we do not reproduce these cases explicitly.

\section{Integrating over many periods}
In Ref.~\cite{Gueron:2006fq} results have been presented from the integration of one period of oscillation. This means on the one hand that the integration time is different for different oscillation frequencies, on the other hand it is not obvious that the effect from the first period can be extrapolated to a longer integration time. As an important result Ref.~\cite{Gueron:2006fq} obtained a characteristic ``plateau'' in the frequency dependence, as can seen in Fig.~\ref{figexpansion}. But since the integration time is not the same for different frequencies, one expect naively that this ``plateau'' represents in fact a linear increase of the effect with the frequency for a fixed integration time.  

\begin{figure}[htb]
	\centering
		\includegraphics[width=8.3cm]{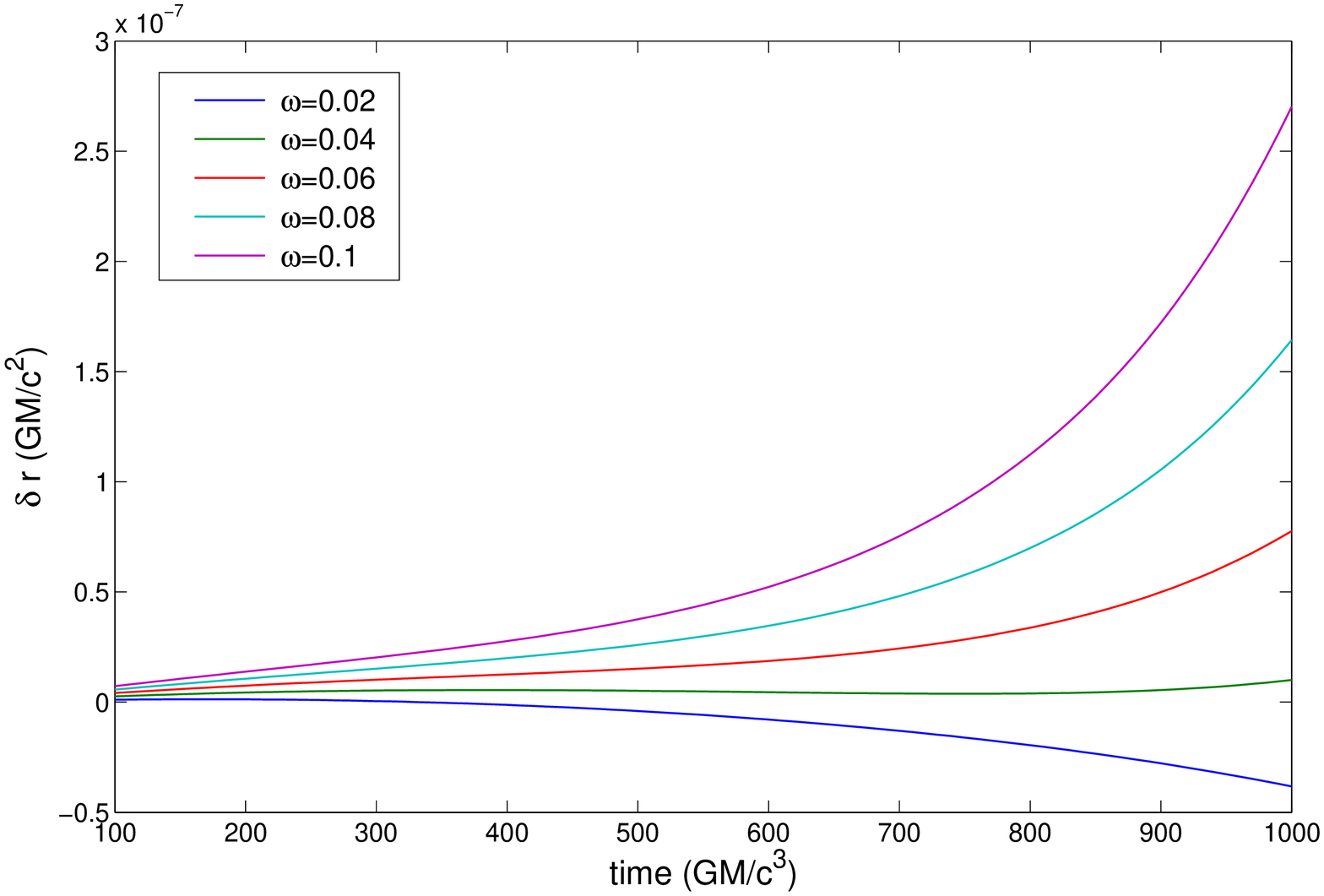}
	\caption{Representation of $\delta r$ as a function of the time (for $R=120$, $\alpha=0.15$ and $\delta_l=5 \ 10^{-3}$) for different values of $\omega$.}
	\label{figTime}
\end{figure}

Thus we present in Figure \ref{figTime} $\delta r$ as a function of time, integrated over many oscillations. It can be seen that the displacement accumulates with time and for relatively small times, the expected linear increase for a fixed frequency is found. However, for low frequencies ($\omega = 0.02$) the positive value of $\delta r$ turns into a negative displecement after some time, which is an effect of strong gravity field. Indeed, more simulations for different $R$ have shown that the effect can increase with the time even for small frequency if $R$ is sufficiently large. 

These results show that for large initial radius $R$ the naive extrapolation of the result of Ref.~\cite{Gueron:2006fq} is correct. From the mentioned ``plateau'' (cf.\ Fig.~\ref{figexpansion}) an integration over a fixed time means that the total displacement grows linearly with the frequency. Thus, it is always interesting to push the frequency in order to enhance the total effect.

\section{Fermi normal coordinates}\label{sec:fermi}

As already mentioned in Ref.~\cite{Gueron:2006fq}, the constraint is implemented in Schwarzschild coordinates and, physically, the strut should be constantly and locally monitored. This could be problematic as it appears to be more obvious to implement the constraint by fabrication or selection of a certain system, whose oscillation characteristics are measured in a laboratory beforehand. During the radial fall experiment an active control of the oscillation should be avoided. In this situation the implementation of the constraint as a periodic motion in Schwarschild coordinates is incorrect. Instead, the constraint should have its prescribed shape in a coordinate system moving together with the vibrating system. While possible in principle, the complexity of this calculation would be beyond the scope of this paper. On the other hand, we could implement the constraint in a locally inertial frame associated to the vibrating system. This is not the simplest method because one need to know \emph{a priori} the trajectory of the vibrating system to describe its associated locally inertial frame. We can also implement the constraint with a good approximation in the locally inertial frame associated to the non-vibrating system. Its trajectory is a free-fall motion~(geodesic), which is non-accelerated (in the sense of General Relativity) and non-rotating (in the sense of Newtonian mechanics). The Fermi normal frame~\cite{Misner:1962} is a realization of such a locally inertial frame.

\begin{figure}[htb]
	\centering
		\includegraphics[width=8.3cm]{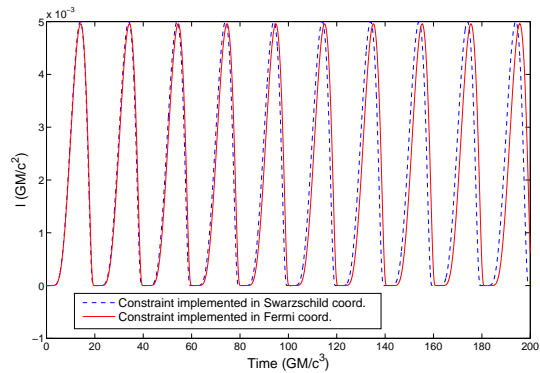}
	\caption{Representation of the constraint expressed in the Schwarzschild coordinates when implemented in Schwarzschild coordinates (dash-dot curve) and when implemented in Fermi normal coordinates (continuous curve). The constraint parameters are $\delta_l=5 \ 10^{-3}$, $\alpha=0.4$, $\omega=0.05$.}
	\label{figConst}
\end{figure}

The general expressions of the coordinate transformations between arbitary coordinates and Fermi normal coordinates can be found in~\cite{Marzlin:1994,Klein:2008}. Using these expressions, we calculate the transformations between the Schwarzschild coordinates and the Fermi normal coordinates associated to the radial free-fall up to fourth order. We implement the constraint~(\ref{CONSTR}) in Fermi normal coordinates and apply the coordinate transformations to obtain the corresponding constraint in Schwarzschild coordinates. The comparison between the two different implementations of the constraint~(\ref{CONSTR}) is shown in Figure~\ref{figConst}. As expected, no essential difference can be seen after one oscillation period. However, the constraint implemented in Fermi normal coordinates fails to be periodic in Schwarschild coordinates and thus after many oscillation periods the frequency between the two constraints is changed. Still, in a weak gravitation and over rather short distances the influence of this shift is negligible.

\section{Conclusions}
In this paper, we made a more detailed analysis of the relativistic glider, originally proposed by Gu\'eron et al.\ \cite{Gueron:2006fq}. We presented an analytical expansion that allows us to confirm the $\frac{1}{R^2}$ behaviour of the effect up to Earth radius. As a consequence, one can expect that it is not possible to see this effect in a very weak gravitational field. On the other hand, we showed that the deviation increases linearly with increasing integration time, which makes a long trajectory interesting for experimental tests.

We performed a more detailed analysis of the dependance of this effect on the initial conditions of the two-body system. If the system is already falling down when it starts to vibrate, the effect will be increased compared to the result of Ref.~\cite{Gueron:2006fq}. On the other hand, if the system is thrown away from the central body, the effect decreases and eventually becomes negative. We also saw that a small tangential velocity has a very small impact on the effect, but it changes completely once the angular velocity is close to the value needed to get a circular orbit around the central body. For a more detailed study of this case we refer to Ref.~\cite{Bergamin:2008ar2}.

Finally, it has been analyzed how the gliding effect changes if the constraint is implemented in the Fermi normal coordinates associated to the reference system instead of Schwarzschild coordinates. As expected, the result does not change considerably in a weak gravitational field. 

\begin{acknowledgements}
 The authors would like to thank D.~Izzo for important discussions on the topic. A.~Hees is research fellow from FRS-FNRS (Belgian Fund for Scientific Research) and he thanks FRS-FNRS for financial support for his thesis at ORB-UCL (Observatoire Royal de Belgique - Universit\'e Catholique de Louvain, Belgium).
\end{acknowledgements}

\bibliographystyle{apsrev}
\bibliography{../../literature}

\end{document}